\documentclass[12pt]{article}
\usepackage{graphicx}
\usepackage{amsmath}

\newcommand{\be}{\begin{equation}}
\newcommand{\ee}{\end{equation}}

\def\emit{\varepsilon}
\def\tth{\tilde{\theta}}
\def\bb{\bar{\beta}}

\begin{document}

\pagestyle{empty}

\renewcommand{\thefootnote}{\fnsymbol{footnote}}


\begin{flushright}
{\small
SLAC--PUB--11781\\
March 2006\\}
\end{flushright}

\vspace{.8cm}

\begin{center}
{\large\bf Emittance Limitation of a Conditioned Beam in a Strong
Focusing FEL Undulator\footnote{Work supported by Department of
Energy contracts DE--AC02--76SF00515.}}

\vspace{1cm}

Z. Huang, G. Stupakov\\
Stanford Linear Accelerator Center,
Stanford, CA  94309\\

\medskip

S. Reiche\\
University of California at Los Angeles, Los Angeles,
CA 90095\\

\end{center}

\vfill

\begin{center}
{\large\bf Abstract }
\end{center}

\begin{quote}

Various methods have been proposed to condition an electron beam
in order to reduce its emittance effect and to improve the
short-wavelength free electron laser (FEL) performance. In this
paper, we show that beam conditioning does not result in a
complete elimination of the emittance effect in an
alternating-gradient focusing FEL undulator. Using a
one-dimensional model and a three-dimensional simulation code, we
derive a criteria for the emittance limitation of a perfectly
conditioned beam that depends on the focusing structure.
\end{quote}

\vfill

\begin{center}
{\it Contributed to the ICFA Beam Dynamics Workshop on \\
the Physics and Applications of High
Brightness Electron Beams \\
Erice, Sicily, Italy (October 9-14, 2005)} \\
\end{center}
\newpage

\section{Introduction}
\label{sec:intro}

A primary factor limiting the performance of short-wavelength,
high-gain free electron lasers (FELs) is the electron beam
transverse emittance. In order to generate transversely coherent
radiation with the diffraction limited emittance $\lambda_r/4\pi$
($\lambda_r$ being the radiation wavelength), the ideal electron
transverse emittance $\emit$ should be less than $\lambda_r/4\pi$
for the efficient beam-radiation interaction. This requirement is
usually not met in the x-ray wavelength region. The spread in the
transverse betatron motion degrades the resonant FEL interaction
and limits the short-wavelength reaches of an x-ray FEL.

To mitigate this problem, it has been proposed to ``condition'' an
electron beam prior to the undulator by increasing each particle's
energy in proportion to the square of its betatron amplitude
\cite{Sessler}. This conditioning enhances the FEL gain by
reducing the axial velocity spread within the electron beam
generated over the undulator, due to both energy spread and finite
transverse emittances.

The original proposal \cite{Sessler} to condition the beam
utilizes a set of cavities in TM$_{210}$ mode immersed in a
focusing lattice. Later approaches to the problem invoke a slow TM
waveguide mode internal to the undulator \cite{Sprangle}, and an
energy chirp in the beam in combination with a chromatic delay
line \cite{Vinokurov,Emma,Emma1,Wolski}. It has also been shown
that an attempt to condition the beam on a short distance is
accompanied by a head-tail focusing variation which can result in
the large effective transverse emittance growth \cite{Emma}.
Although this emittance growth can be avoided \cite{Wolski}, the
pace of conditioning becomes much slower, and the required length
of the conditioner considerably increases. Other proposals are
based on using lasers and include Thomson backscattering
\cite{Schroeder}, and interaction of the laser with the electron
beam in two dedicated undulators \cite{Zholents}. Recently, beam
conditioning using nonlinear RF acceleration is also discussed
\cite{stupakov}.

In the FEL gain analysis of the original proposal \cite{Sessler},
the undulator natural focusing is assumed to confine the electron
beam in both transverse directions. This results in the complete
elimination of the emittance effect for a perfectly conditioned
beam. In this paper, we study a more practical situation when the
transverse focusing is provided by alternating-gradient
quadrupoles in the undulator, as found in typical x-ray FEL
designs. We show that although conditioning eliminates the effect
of the average slippage of electrons relative to the radiation
phase, there remain phase oscillations with the period given by
the focusing lattice. These phase oscillations can affect the FEL
performance and impose a limitation on the beam emittance even for
a perfectly conditioned beam, albeit this emittance limitation is
much relaxed as compared to a unconditioned beam.

\section{Emittance effect in a strong focusing undulator}

Although a magnetic undulator can provide natural focusing in both
transverse planes~\cite{scharlemann85}, the focusing strength is
typically too weak for the high-energy electron beam that drives
an x-ray FEL. Thus, alternating-gradient quadrupole magnets are
inserted in undulator breaks to provide the necessary strong
focusing, usually in the form of a FODO lattice (consisting of
repetitive focusing-undulator-defocusing-undulator cells). The
horizontal betatron motion is given by
    \begin{align}
    x_\beta (z) =& \sqrt{2J_x\beta_x} \cos\Phi_x(z)\,, \nonumber \\
    p_x (z) \equiv & {dx_\beta\over dz}= -\sqrt{2J_x\over \beta_x} \left[\sin\Phi_x(z) +
    \alpha_x\cos\Phi_x(z)
    \right]\,,  \label{eq:px}
    \end{align}
where $J_x$ is the horizontal action of the electron, $\alpha_x$
and $\beta_x$ are the Twiss parameters, and $p_x$ denotes the
angle of the orbit with the $z$ axis. The second term in
Eq.~(\ref{eq:px}) is ignored in Ref.~\cite{yu95} under the smooth
approximation. Its importance in FEL dynamics and emittance
compensation (i.e., conditioning) is pointed out in
Ref.~\cite{reiche}.

\begin{figure}
\centering
    \includegraphics[width=0.6\linewidth]{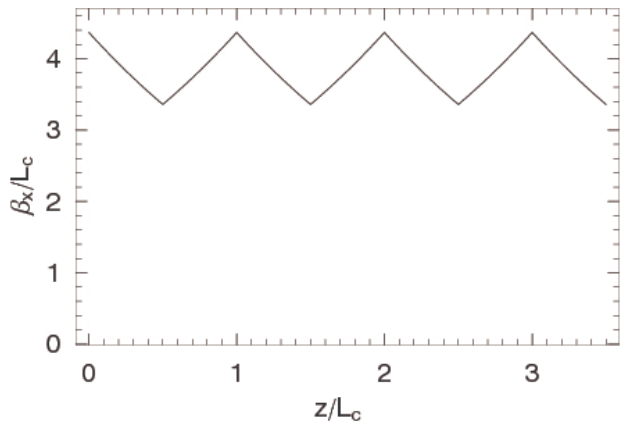}
    \caption{Variation of the beta function along the distance of the lattice
    for the phase advance per cell of 15 degrees. The derivative $d\beta_x/dz$ is
    close to the values $\pm 2$, but the deviation of $\beta_x$ from the average value
    $\bb$ is relatively small.}
    \label{fig:FODO}
\end{figure}

To avoid a large beam size variation in the undulator, the FODO
lattice is usually designed to have a small phase advance per
cell. In this case, the average betatron wavelength $2\pi \bb$ is
much larger than the FODO cell length $L_c$. In a first
approximation the beta function is constant over the length of the
undulator, and the accumulated betatron phase is
    \be
    \Phi_x (z) = \phi_x+\int_0^z {ds\over \beta_x(s)} \approx
    \phi_x+{z\over \bb}\,,
    \ee
where $\phi_x$ is the initial electron phase. More importantly for
the emittance compensation, such a FODO lattice has~\cite{reiche}
    \be
    \alpha_x=-{1\over 2}{d\beta_x\over dz}\approx \pm 1
    +O\left({L_c^2\over \bb^2}\right)
    \ee
with the sign alternating per half cell. Fig. \ref{fig:FODO}
illustrates the variation of the beta function in a FODO lattice
with 15 degrees phase advance per cell.

Under this short-cell-length approximation, the square of the
horizontal divergence is
    \be
    p_x^2 \approx {2J_x\over \bb} \left[1\pm \sin\left({2z\over \bb} +2\phi_x
    \right)\right]\,.
    \ee
Similarly in the vertical $y$ direction, we have
    \be
    p_y^2 \approx {2J_y\over \bb} \left[1\mp \sin\left({2z\over \bb}
    +2\phi_y
    \right)\right]\,.
    \ee
Here the signs $\pm$ becomes $\mp$ due to the focusing-defocusing
asymmetry. We have also assumed that the average beta function is
the same in both directions.

The FEL interaction is a resonant phenomenon that depends
critically on the evolution of the electron phase relative to the
co-propagating radiation field. The so-called ponderomotive phase
is defined as $\theta=(k_r+k_u)z-ck_r\bar{t}$, where
$k_r=2\pi/\lambda_r$ is the radiation wavenumber,
$k_u=2\pi/\lambda_u$, $\lambda_u$ is the undulator period, and
$\bar{t}$ is the electron arrival time at the location $z$
averaged over the undulator period. The rate of the phase change
is
    \be
    {d\theta\over dz} = (k_r+k_u) - {ck_r\over v_z}\,,
    \label{eq:phase}
    \ee
where $v_z$ is the undulator-period-averaged longitudinal velocity
and is given by
    \be
    {v_z\over c} \approx 1-{1\over 2\gamma^2}-{v_\perp^2\over 2}
    = 1-{1+K^2/2 \over 2\gamma^2} - {p_x^2+p_y^2 \over 2}\,.
    \ee
Here $K$ is the undulator parameter and is approximately constant.
Its weak dependence on $x$ and $y$ can be neglected as long as the
external focusing of the FODO lattice is much stronger than the
natural focusing of the undulator. Making use of the resonant
condition
    \be
    k_r={2\gamma_r^2 k_u\over 1+K^2/2} \,,
    \ee
we can write Eq.~(\ref{eq:phase}) as
    \begin{align}
    {d\theta\over dz} =& 2k_u {\Delta \gamma\over \gamma_r}- {k_r\over 2}
    (p_x^2+p_y^2)\nonumber \\
    = & 2k_u {\Delta \gamma\over \gamma_r}- k_r {(J_x+J_y)\over
    \bb} \nonumber \\
    - & k_r \left[
    \pm {J_x\over \beta} \sin\left({2z\over \bb} +2\phi_x\right)
    \mp {J_y\over \beta} \sin\left({2z\over \bb}
    +2\phi_y\right)\right]\,.
    \label{eq:phase1}
    \end{align}

Since the last term of Eq.~(\ref{eq:phase1}) is oscillatory with
the FODO lattice period (cell length), the main accumulating
effects on the ponderomotive phase are the first two terms.
Neglecting this oscillatory term, the phase equation in such a
FODO cell is identical to that in a natural focusing
undulator~\cite{note}, as noted in Refs.~\cite{reiche,saldin01}.
For a beam with finite energy spread and emittance, the first two
terms introduce phase slippage of the electron relative to the
radiation and result in the phase spread of the beam. Denoting
$\langle\rangle$ as the average over the beam, we have $\langle
J_x\rangle =\emit_x$ and $\langle J_y\rangle =\emit_y$, where
$\emit_{x,y}=\emit_n/\gamma_r$ are the beam transverse emittances.
In order to not significantly degrade the FEL performance, the
emittance-induced phase spread over one FEL power gain length
$L_G$ should be less than unity, i.e.,
    \be
    {2 k_r\emit_n\over \gamma_r \bb}L_G < 1\,, \quad \text{or} \quad
    \emit_n <\gamma_r {\lambda_r \bb \over 4\pi L_G}\,.
    \label{eq:uncond}
    \ee

This situation changes if the initial electron energy can be
conditioned to its transverse betatron amplitudes such that
    \be
    2k_u {\Delta \gamma_0\over \gamma_r} = k_r {(J_x+J_y)\over
    \bb} \,,
    \ee
the dominant emittance effect (i.e., the second term in
Eq.~(\ref{eq:phase1})) is then removed from Eq.~(\ref{eq:phase1}).
Averaging over $J_x$ and $J_y$, the (transversely) correlated
energy spread required to condition a beam with the normalized
emittance $\gamma_r\emit_x=\gamma_r\emit_y=\emit_n$ is
    \be
    \langle \Delta \gamma_0\rangle_c = {\lambda_u\over \lambda_r}
    {\emit_n\over \bb}\,. \label{eq:cond}
    \ee

However, when the beam is perfectly conditioned to satisfied
Eq.~(\ref{eq:cond}), the last oscillatory term in
Eq.~(\ref{eq:phase1}) is no longer negligible and can play a
limiting role with a large enough emittance. We discuss
quantitatively this remaining emittance effect on the FEL
performance for a conditioned beam in the following two sections.

\section{1D model of the ponderomotive phase oscillation}

The ponderomotive phase oscillations in Eq.~(\ref{eq:phase1})
depend on the transverse variables through $J_x$, $J_y$, $\phi_x$
and $\phi_y$ and is a three-dimensional (3D) problem. To isolate
the oscillation effect and to simplify the problem, we study a
heuristic one-dimensional (1D) model with the following phase
equation
    \be
    {d\theta\over dz}=2 k_u \delta +{k_r \emit_n\over \gamma_r\bb} f(z)\,,
    \label{eq:phase1D}
    \ee
where $\delta=(\Delta\gamma-\Delta\gamma_0)/\gamma_r$ is the
FEL-induced energy change, and $f(z)$ represents the oscillatory
behavior introduced by a conditioned beam in the FODO lattice
(with a cell length much smaller than the betatron wavelength),
i.e.,
    \be
    f(z) = \begin{cases}
      +1 & \text{when $(n-1)L_c\le z<(2n-1){L_c\over 2}$}\,,  \\
      -1 & \text{when $(2n-1){L_c\over 2}\le z<nL_c$}\,,
    \end{cases}
    \ee
and $n=1,2,3...$. In the absence of the FEL interaction (i.e.,
when $\delta=0$), the phase oscillates between 0 and $\theta_0$,
with the maximum phase deviation
    \be
    \theta_0 = {k_r \emit_n L_c\over 2\gamma_r\bb}\,.
    \ee

Using the FEL Pierce parameter $\rho$~\cite{BPN}, we introduce a
scaled distance $\tau=2k_u\rho z$ and a scaled energy
$\eta=\delta/\rho$. Equation~(\ref{eq:phase1D}) becomes
    \be
    {d\theta\over d\tau} = \eta + \xi\,, \quad \text{with}
    \quad \xi={k_r \emit_n\over 2\gamma_rk_u\rho\bb} f(\tau)\,.
    \label{eq:scaled1}
    \ee

The FEL-induced energy change is
    \be
    {d\eta\over d\tau} = a e^{i\theta} +\text{complex
    conjugate}\,,  \label{eq:scaled2}
    \ee
where $a$ is the slowly-varying radiation field amplitude
(properly scaled by $\rho$). Neglecting any transverse dependence
and considering the radiation field at the resonant frequency
$ck_r$, the 1D field equation is
    \be
    {da\over d\tau} = -\langle e^{-i\theta}\rangle \,.  \label{eq:scaled3}
    \ee

This set of coupled equation can be solved by averaging over the
fast oscillation if its period is much smaller than the field gain
length, similar to the undulator-period averaging procedure for a
planar undulator. To illustrate this process, we define
    \be
    \tth=\theta -\int^\tau \xi(\tau')d\tau'\,,
    \ee
Equation~(\ref{eq:scaled1}) can be written as
    \be
    {d\tth\over d\tau} =\eta\,.\label{eq:scaled4}
    \ee
Equations~(\ref{eq:scaled2}) and ~(\ref{eq:scaled3}) become
    \begin{align}
    {d\eta\over d\tau} =& a e^{i\tth} \exp\left(i\int_0^\tau \xi(\tau')d\tau'\right) +\text{complex
    conjugate}\,, \label{eq:scaled5} \\
    {da\over d\tau} =& -\langle e^{-i\tth}\rangle \exp\left(-i\int_0^\tau \xi(\tau')d\tau'\right)\,.
    \label{eq:scaled6}
    \end{align}
Treating $\eta$, $\tth$, and $a$ as slowly-varying variables, we
can average Eqs.~(\ref{eq:scaled5}) and (\ref{eq:scaled6}) over
the oscillation period $L_c$ to obtain
    \begin{align}
    {d\eta\over d\tau} =& a e^{i\tth} A(\theta_0) +\text{complex
    conjugate}\,,  \label{eq:average1} \\
    \bar{da\over d\tau} =& -\langle e^{-i\tth}\rangle
    A(-\theta_0)\,, \label{eq:average2}
    \end{align}
where
    \begin{align}
    A(\theta_0) &=\int_0^{L_c} {dz\over L_c} \exp\left(i\int_0^\tau
    \xi(\tau')d\tau'\right)\nonumber\\
     =& \int_0^{L_c} {dz\over L_c}
    \exp\left(i\theta_0 \int_0^{z} {2ds\over L_c}
    f(s) \right) \,.
    \end{align}

\begin{figure}
\centering
    \includegraphics[width=0.6\linewidth]{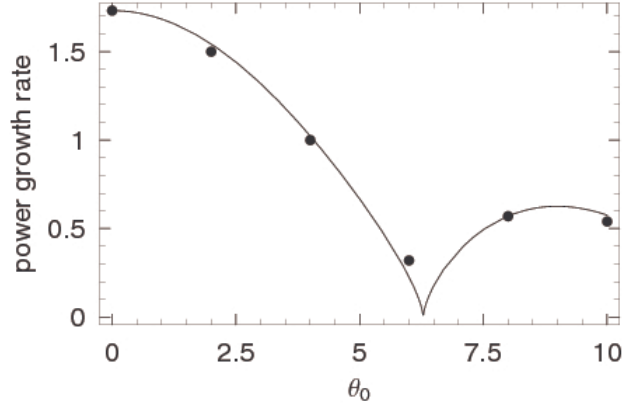}
    \caption{1D FEL power growth rate under a rapid phase oscillation with the maximum phase deviation
    $\theta_0$. Solid curve is the solution of Eq.~\ref{eq:cubic}, and the symbols represent 1D simulation results.}
    \label{fig:1d}
\end{figure}

Equations~(\ref{eq:scaled4}), (\ref{eq:average1}) and
(\ref{eq:average2}) are the FODO-cell averaged FEL equations and
can be solved with the usual techniques (see, e.g.,
Ref.~\cite{BPN}). Assuming that $a(\tau) \propto e^{-i\mu\tau}$,
we obtain a cubic equation for the complex growth rate $\mu$:
    \be
    \mu^3 = \vert A(\theta_0)\vert^2\,. \label{eq:cubic}
    \ee
The FEL power growth rate 2Im$\mu=\sqrt{3}\vert
A(\theta_0)\vert^{2/3}$ versus the maximum phase deviation
$\theta_0$ is shown in Fig.~\ref{fig:1d}. It agrees well with the
1D simulation results by solving the original set of
Eqs.~(\ref{eq:scaled1}),~(\ref{eq:scaled2}) and
~(\ref{eq:scaled3}) for an electron beam without any initial
energy spread. The FEL growth rate is degraded when
    \be
    \theta_0 = {k_r \emit_n L\over 2\gamma_r\beta}> 1\,.
    \ee

\section{3D GENESIS simulations}

The above 1D model shows that when the maximum phase oscillation
amplitude exceeds unity, the FEL gain will be degraded as compared
to the ideal case. Therefore, for a perfectly conditioned beam
that satisfies Eq.~(\ref{eq:cond}) in a FODO lattice, the FEL
performance will still be affected by the emittance-induced phase
oscillations when the normalized emittance $\emit_n$ exceeds a
critical emittance given by
    \be
    \emit_n^c \equiv \gamma_r {\lambda_r \bb\over \pi L_c}\,,
    \label{eq:Cemit}
    \ee

\begin{figure}
\centering
    \includegraphics[width=0.6\linewidth]{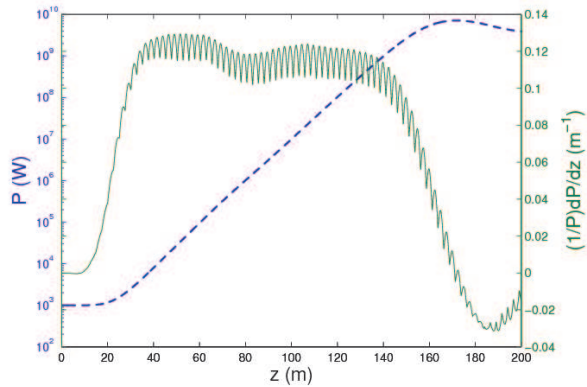}
    \caption{(Color) GENESIS amplifier run of a conditioned beam for $L_c=5.04$~m, $\bb=19$~m and $\emit_n=10$~$\mu$m.
    The blue dashed curve is the FEL power $P$ at $\lambda_r=$1 \AA, while the green solid curve is the local growth rate $1/P(dP/dz)$.}
    \label{fig:genesis}
\end{figure}

Since this emittance criteria is derived with a heuristic 1D
model, we examine it using the 3D FEL code GENESIS~\cite{genesis}.
The electron beam and undulator parameters are given in
Table~\ref{tab:para}. A typical FEL power evolution ($P$ versus
$z$) is shown in Fig.~\ref{fig:genesis} (blue dashed curve). The
local power growth rate $1/P(dP/dz)$ is also plotted in the same
figure (green solid curve) and is oscillatory with a period
equaling to $L_c/2$ as a result of emittance-induced phase
oscillations.

\begin{table}
\begin{center}
\caption{\label{tab:para} GENESIS FEL simulation parameters of
conditioned beams in a strong focusing undulator.}
\begin{tabular}{|llc|}
\hline Parameter & Symbol & Value \\
\hline electron energy & $\gamma_r mc^2$ & 20 GeV
\\ flattop bunch current & $I_{pk}$ & 3 kA
\\ transverse norm. emittance & $\emit_n$ & 0.1 to 19~$\mu$m
\\ conditioned energy spread & $\langle\Delta \gamma_0\rangle_c$ & given by
Eq.~(\ref{eq:cond})
\\ uncorrelated energy spread & $\sigma_{\gamma_0}$ & 0
\\ undulator period & $\lambda_u$ & 3~cm
\\ undulator parameter & $K$ & 4.67
\\ average beta function & $\bb$ & 4.8/9.4/19~m
\\ FODO cell length & $L_c$ & 2.64/5.04/9.83~m
\\ FEL wavelength & $\lambda_r$ & 1.0~\AA
\\ \hline
\end{tabular}
\end{center}
\end{table}

\begin{figure}
\centering
    \includegraphics[width=0.6\linewidth]{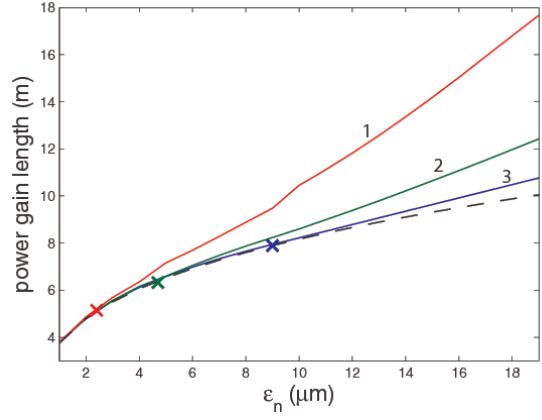}
    \caption{(Color) FEL power gain length from GENESIS simulations at $\bb=19$~m for $L_c=2.64$~m
    (blue solid curve 1), $L_c=5.04$~m (green solid curve 2),
    $L_c=9.83$~m (red solid curve 3), and from 3D theory without any phase oscillation (black dashed curve).
    The colored symbols represent the corresponding critical emittance for each lattice (determined
    by Eq.~(\ref{eq:Cemit})) when the phase oscillation is expected to increase the gain
length.}
    \label{fig:beta20m}
\end{figure}

We extract the FEL power gain length by averaging a relatively
constant $1/P(dP/dz)$ over many oscillation periods in the GENESIS
simulations. The resulting gain length versus the normalized
emittance is plotted in Fig.~\ref{fig:beta20m} for $\bb=19$~m and
for three different FODO cell lengths. We also compare the
simulation results with the theoretical gain length for a
conditioned beam in the absence of any phase oscillation. The
latter is obtained by solving the usual 3D eigenmode equation
(see, e.g., Ref~\cite{xie2000}) for the fundamental Gaussian mode
without any energy and angular spreads. As shown in
Fig.~\ref{fig:beta20m}, the simulation results agree with the 3D
theory that does not take into account phase oscillations up to
the critical emittances predicted by Eq.~(\ref{eq:Cemit}). For
emittances exceed these critical values, the gain lengths
extracted from simulations start to increase faster than the ideal
case.

\begin{figure}
\centering
    \includegraphics[width=0.6\linewidth]{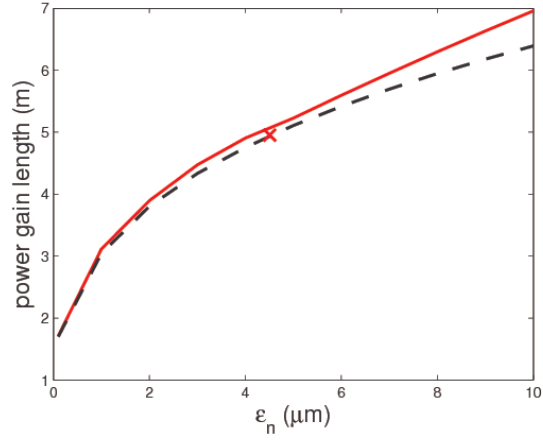}
    \caption{(Color) FEL power gain length from GENESIS simulations at $\bb=9.4$~m for $L_c=2.64$~m (red solid curve),
    and from 3D theory without any phase oscillation (black dashed curve). The symbol
    represents the corresponding critical emittance (determined by Eq.~(\ref{eq:Cemit}))
    when the phase oscillation is expected to increase the gain length.}
    \label{fig:beta10m}
\end{figure}

\begin{figure}
\centering
    \includegraphics[width=0.6\linewidth]{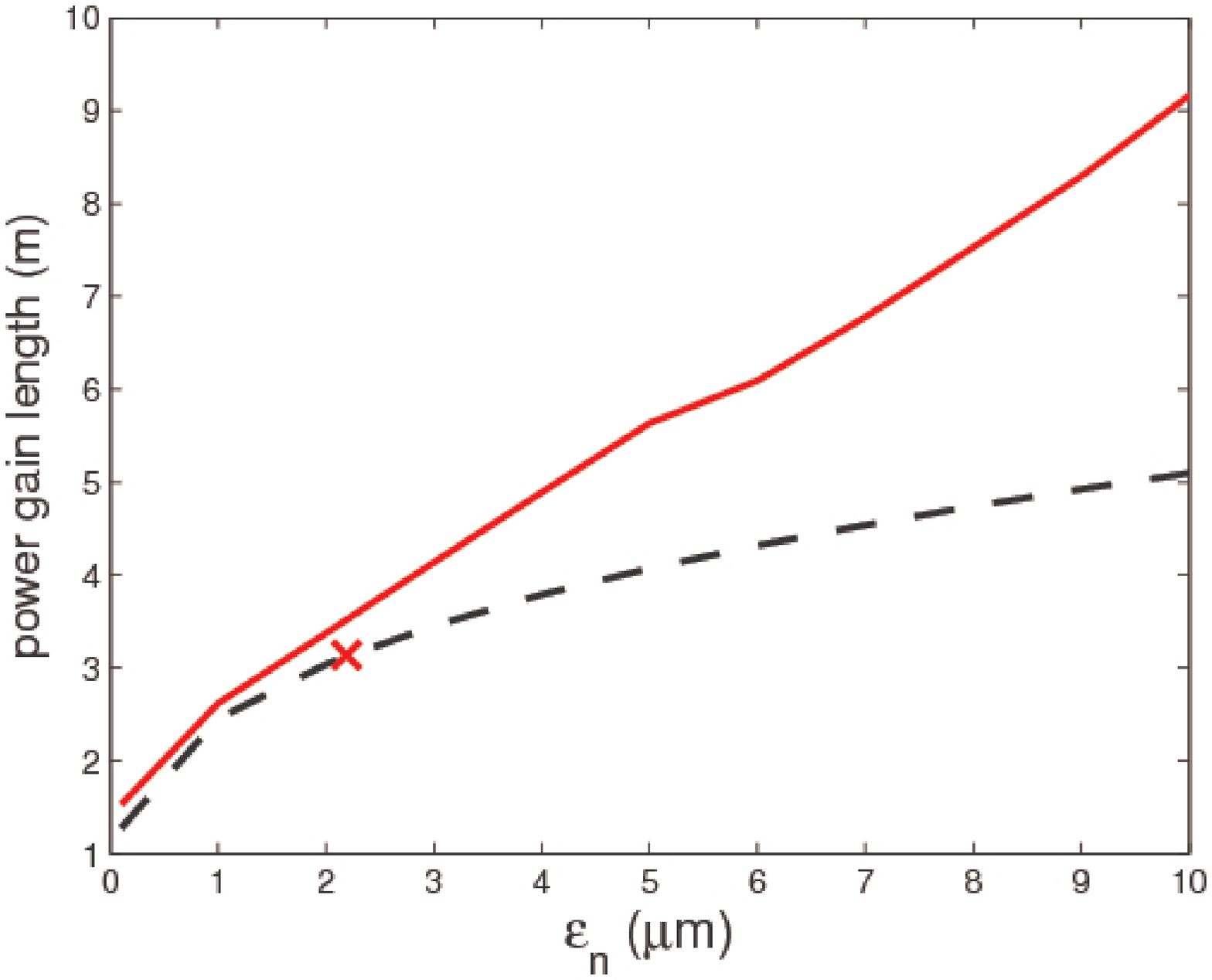}
    \caption{(Color) FEL power gain length from GENESIS simulations at $\bb=4.8$~m for $L_c=2.64$~m (red solid curve),
    and from 3D theory without any phase oscillation (black dashed curve). The symbol
    represents the corresponding critical emittance (determined by Eq.~(\ref{eq:Cemit}))
    when the phase oscillation is expected to increase the gain length.}
    \label{fig:beta5m}
\end{figure}

We have also performed GENESIS simulations by varying the average
beta function $\bb$ while keeping the same focusing structure
(with $L_c=2.64$~m). Figures~\ref{fig:beta10m} and
\ref{fig:beta5m} show the gain length comparisons for $L_c=2.64$~m
at $\bb=9.4$~m and 4.8~m, respectively. Combining with
Fig.~\ref{fig:beta20m}, we see that Eq.~(\ref{eq:Cemit}) is
reasonably accurate in predicting the critical emittances for
different beta functions and FODO lattices.

As the ratio of the average beta function to the cell length
becomes too large (e.g., when $\bb=39$~m and $L_c=2.64$~m or when
$\bb=19$~m and $L_c=1.44$~m), the gain length extracted from the
GENESIS simulation starts to deviate from the 3D theory (without
any phase oscillation) at a smaller emittance than that predicted
by Eq.~(\ref{eq:Cemit}). A plausible explanation is given as
follows. In x-ray FELs the transverse coherence builds up slowly
as many higher-order transverse modes are excited by an electron
beam with a relatively large emittance. To some degrees a periodic
modulation in the electron beam size along the undulator is
beneficial to the build-up of the transverse coherence (namely
higher order modes are suppressed by a varying beam
envelope~\cite{reicheThesis}). When the ratio of the beta function
to the cell length becomes too large, this higher-order
suppression mechanism is not effective since the beam size is
almost constant, and the growth of the radiation is contributed
from both the fundamental mode as well as higher-order modes. In
this case, the comparison between simulations (including
higher-order modes) and the theory (for the fundamental mode) is
no longer valid.

\section{Summary}

In this paper, we study the FEL performance for a perfectly
conditioned beam in a strong focusing undulator consisting of FODO
cells. We develop a heuristic 1D model for the emittance-induced
phase oscillation and show that the maximum phase deviation should
be less than unity in order not to degrade the FEL gain. This
criteria limits the maximum emittance that may be conditioned in
such a focusing channel and is confirmed by the 3D GENESIS
simulations over a reasonably wide parameter range. Therefore,
instead of the usual emittance criteria for the unconditioned beam
as given by Eq.~(\ref{eq:uncond}), the emittance of the
conditioned beam in a FODO lattice with an average beta function
$\bb$ and a cell length $L_c$ must satisfy
    \be
    \emit_n < \gamma_r {\lambda_r\bb\over \pi L_c}\,.
    \ee
That is, the emittance requirement for a conditioned beam is much
relaxed but not without any limitation.

\section{Acknowledgments}

We thank A. Zholents for useful discussions. This work was
supported by Department of Energy contract DE--AC02--76SF00515.


\begin{thebibliography}{1}
\bibitem{note}
In a natural focusing undulator, $\alpha_{x,y}=0$ in
Eq.~(\ref{eq:px}), but the transverse variations of the undulator
parameter $K$ together with the first term in the bracket of
Eq.~(\ref{eq:px}) result in the phase equation ${d\theta/dz}=2k_u
{\Delta \gamma/ \gamma_r}- k_r {(J_x+J_y)/\beta_n}$, where
$\beta_n=\gamma_r\lambda_u/(\pi K)$ is the natural focusing beta
function. See also Ref.~\cite{scharlemann85}.
\end{thebibliography}
\end{document}